# Semi-supervised Embedding in Attributed Networks with Outliers

Jiongqian Liang*†     Peter Jacobs*     Jiankai Sun*     Srinivasan Parthasarathy*†

**Abstract**

In this paper, we propose a novel framework, called Semi-supervised Embedding in Attributed Networks with Outliers (SEANO), to learn a low-dimensional vector representation that systematically captures the topological proximity, attribute affinity and label similarity of vertices in a *partially labeled attributed network* (PLAN). Our method is designed to work in both transductive and inductive settings while explicitly alleviating noise effects from outliers. Experimental results on various datasets drawn from the web, text and image domains demonstrate the advantages of SEANO over the state-of-the-art methods in semi-supervised classification under transductive as well as inductive settings. We also show that a subset of parameters in SEANO are interpretable as outlier scores and can significantly outperform baseline methods when applied for detecting network outliers. Finally, we present the use of SEANO in a challenging real-world setting – flood mapping of satellite images and show that it is able to outperform modern remote sensing algorithms for this task.

## 1 Introduction

Many applications are modeled and analyzed as attributed networks, where vertices represent entities with attributes and edges express the interactions or relationships between entities. In many scenarios, one also has knowledge about the labels of some vertices in an attributed network. Such networks are referred to as partially labeled attributed networks (PLANs). While PLANs contain much richer information than plain networks, they are also more challenging to analyze. In view of the tremendous success of *network embedding* [34, 27, 9] in plain networks for graph mining tasks such as vertex classification [38] and network visualization [34], researchers have adapted similar ideas to attributed networks [13, 37, 26, 15]. However, there are three key challenges on node embedding in a PLAN:

1. How to learn the network embeddings by collectively incorporating heterogeneous information in the PLAN, including graph structure, vertex attributes, and **partially** available labels?
2. How to perform inductive embedding learning so that embeddings can be generated for vertices unobserved during the training phase?
3. How to address the outliers in the PLAN and learn more robust embeddings in a noisy environment?

Several efforts have focused on the first challenge – to capture topological structure, vertex attributes and label information for embedding learning in a transductive manner (assuming all the vertices are accessible during training) [26, 13, 37, 15, 10], although some of them cannot be easily adapted to a semi-supervised learning setting [13, 10]. Solutions in the inductive setting for PLANs, which support generating embeddings for unseen nodes, have only been looked at very recently [37, 10]. To the best of our knowledge, there is no previous work explicitly accounting for the effect of outliers in network embedding. In this paper, we propose a novel approach to simultaneously overcome these three challenges.

Specifically, we design a dual-input and dual-output deep neural network to inductively learn vertex embeddings. The input layers are hinged on the attributes of the vertices and their neighborhoods respectively while the output layers provide label and context predictions. The two different output layers respectively form the supervised and unsupervised components of our model. By alternately training the two components on the PLAN, we *learn a unified embedding encompassing information related to structure, attributes, and labels*. We also show that our method can generate quality embeddings for new vertices unseen during training, therefore supporting inductive learning in nature. Furthermore, our model explicitly accounts for the notion of outliers during training and is capable of effectively alleviating the potentially adverse impact from the anomalous vertices on the learned embeddings. We also reveal a nice property of the proposed model that **a particular set of parameters can be interpreted as outlier scores for the vertices in the PLAN**. We refer to our method as <u>S</u>emi-supervised <u>E</u>mbedding in <u>A</u>ttributed <u>N</u>etworks with <u>O</u>utliers (SEANO). We empirically evaluate the quality of the embeddings generated by SEANO through semi-supervised classification and show that SEANO significantly outperforms state-of-the-art methods in both transductive and inductive settings. In addition, we conduct outlier detection based on the output outlier scores from `SEANO` and demonstrate its advantages over baseline methods specializing in network outlier detection. We finally conduct a case study of flood mapping to visually show the power of `SEANO` when applied to flood mapping.

*The Ohio State University.
†Primary contact: {liangji, srini}@cse.ohio-state.edu.

## 2 Related Work

**Network Embedding**: Network embedding strategies have gained increasing importance in recent years. Early ideas include IsoMap [35] and Locally Linear Embedding (LLE) [29], which exploited the manifold structure of vector data to compute low-dimensional embeddings. More recently, due to the emergence of naturally arising network data, other network embedding methods have been proposed [34, 27, 9]. In addition to learning embeddings for homogeneous networks, several researchers have proposed ideas for embedding attributed networks [36, 13, 12, 26, 37, 15, 10]. While they incorporate the attributes and/or label information into the embeddings, most are inherently transductive and cannot generate embeddings for vertices unseen during training. The two exceptions are Planetoid [37] and GraphSAGE [10] for inductive learning. However, Planetoid [37] is specialized for semi-supervised classification and the output network embedding, as a byproduct, does not capture all the information (as can be seen from their model architecture). Therefore, its embeddings might not be generalized to other applications such as visualization and clustering. GraphSAGE [10], on the other hand, only works on unsupervised learning or fully supervised learning setting and cannot be directly applied in a semi-supervised manner. Finally, none of the existing work on network embedding explicitly accounts for the impact of outliers. We summarize the differences between the proposed `SEANO` model with some of these recent efforts in Table 1.

| Method | Attributes | Labels | Semi-supervised | Inductive | Address Outliers |
|---|---|---|---|---|---|
| DeepWalk [27] | ✗ | ✗ | ✗ | ✗ | ✗ |
| Node2Vec [9] | ✗ | ✗ | ✗ | ✗ | ✗ |
| TADW [36] | ✓ | ✗ | ✗ | ✗ | ✗ |
| LANE [13] | ✓ | ✓ | ✗ | ✗ | ✗ |
| TriDNR [26] | ✓ | ✓ | ✓ | ✗ | ✗ |
| Planetoid [37] | ✓ | ✓ | ✓ | ✓ | ✗ |
| GCN [15] | ✓ | ✓ | ✓ | ✗ | ✗ |
| GraphSAGE [10] | ✓ | ✓ | ✗ | ✓ | ✗ |
| SEANO | ✓ | ✓ | ✓ | ✓ | ✓ |

Table 1: A comparison of `SEANO` with baseline methods. ✓ means using a certain type of information or support certain functionality while ✗ indicates the opposite.

**Outlier Detection:** While there has been a plethora of work on outlier detection under different contexts [3, 19], outlier detection in network data has not been studied until recent years [1]. Previous methods in this space mostly focused on the topological features of the graph to detect anomalous patterns, such as subgraph frequency [24], community structure [7], etc. More recent work in this area looks into the attributed network by incorporating the vertex attributes [28, 21, 17]. Only a couple of recent works attempted to discover network outliers using network embeddings [6, 11]. These efforts, however, do not apply to attributed networks. Our work is somewhat orthogonal to these efforts as we shall discuss shortly.

## 3 Methodology

**3.1 Problem Formulation** We first define a Partially Labeled Attributed Network (PLAN) and formulate our framework for semi-supervised embedding on attributed networks with outliers (`SEANO`).

DEFINITION 1. *Partially Labeled Attributed Network. A partially labeled attributed network is an undirected graph $\mathcal{G} = (V, E, X, Y)$, where: $V = \{1, 2, ..., n\}$ is the set of vertices; $E$ is the set of edges; $X = (\mathbf{x}_1, \mathbf{x}_2, ..., \mathbf{x}_n)$ is the attribute information matrix; and $Y = (y_1, y_2, ..., y_n)$ are the labels of the vertices in $V$, most of which are unknown.*

Depending on whether the label of a vertex is known, we divide the vertices into labeled vertices $V_L$ and unlabeled vertices $V_U$. We also consider the potentially negative effect of *network outliers*. Following the definition elsewhere [21], we define *network outliers* in a PLAN as the vertices whose attributes significantly deviate from the underlying attributes distribution of their context localized by the graph structure and vertex label. An example outlier in a PLAN would be a vertex containing very different attributes from other vertices in a densely connected component with the same label. With the concept of the PLAN and the network outlier, we define our embedding learning problem as follows.

DEFINITION 2. *Semi-supervised Embedding in Attributed Networks with Outliers. Given a partially labeled attributed network $\mathcal{G} = (V, E, X, Y)$ with a small portion of unknown network outliers, we aim to learn a robust low-dimensional vector representation $\mathbf{e}_i \in \mathbb{R}^r$ for each vertex $i$, where $r \ll n$, s.t. $\mathbf{e}_i$ can jointly capture the information of the attributes, graph structure, and the partial labels in the PLAN.*

**3.2 The Proposed Model** We propose `SEANO` to solve the above problem. The architecture of `SEANO` (as illustrated in Figure 1) consists of a deep model with two input layers and two output layers. The inputs and outputs are connected through a series of non-linear mapping functions, which transform the features into a non-linear latent space.

As shown in Figure 1, the two input layers are the attributes of the vertex $\mathbf{x}_i$ and the average attributes of its neighborhoods $\overline{\mathbf{x}}_{N_i}$[1]. They go through the **same set** of non-linear mapping functions ($l_1$ layers) and are aggregated in the embedding layer through a weighted sum. The two output layers are hinged on the embedding layer. The left output layer in Figure 1 predicts the class label $y_i$ of the input vertex while the right output layer yields the context of the network input.

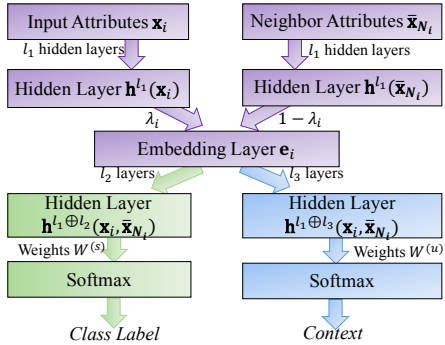

Figure 1: Network architecture: the feed-forward neural network with dual inputs and dual outputs.

| Symbol | Definition |
|---|---|
| $\mathcal{G}, V, E$ | partially labeled attributed network, vertex set, edge set |
| $X, Y$ | attribute information matrix, vertex labels |
| $n, d, r$ | # vertices, # attributes, # embedding dimensions |
| $V_L, V_U$ | labeled vertices, unlabeled vertices |
| $N_i$ | neighborhoods of vertex $i$ |
| $\mathbf{x}_i$ | attributes of vertex $i$ |
| $\overline{\mathbf{x}}_{N_i}$ | avg. attributes of vertex $i$'s neighborhoods $N_i$ |
| $\lambda_i, \mathbf{e}_i$ | aggregation weight of vertex $i$, embedding of vertex $i$ |
| $\mathbf{h}^k(\cdot)$ | values at the $k$-th layer |
| $\mathbf{W}^{(s)}, \mathbf{W}^{(u)}$ | weights at the (un)supervised output layer |
| $L_s, L_u$ | loss function of (un)supervised learning |

Table 2: Table of notations

**3.2.1 Architecture and Rationale** We denote the $k$-th layer of the deep model as $\mathbf{h}^k(\mathbf{x}_i) = \phi\left(\mathbf{W}^k \mathbf{h}^{k-1}(\mathbf{x}_i) + \mathbf{b}^k\right)$, where $\mathbf{W}^k$ and $\mathbf{b}^k$ are the weights and biases in the $k$-th layer, and $\phi(\cdot)$ is a non-linear activation function[2]. Then the embedding layer can be represented as

$$(3.1) \qquad \mathbf{e}_i = \lambda_i \mathbf{h}^{l_1}(\mathbf{x}_i) + (1 - \lambda_i) \mathbf{h}^{l_1}(\overline{\mathbf{x}}_{N_i})$$

where $\lambda_i \in [0, 1]$ is a parameter associated with each vertex $i$ (called *aggregation weight*) and is learned through the model training[3]. We now illustrate how the proposed model can be used to address the three aforementioned challenges below.

**Semi-supervised Embedding Learning:** The left output layer is considered as the supervised learning component of the model since it predicts the class label and we use labeled data to train it. The right output layer predicts the context of the network input, which comes from the context generation algorithm, such as random walks [27] (see Sec. 3.2.2). Therefore, this part is regarded as the unsupervised component of the model that captures the topological structure information. These two parts are tightly inter-connected as they share the first $l_1 + 1$ layers from input layers to the embedding layer. Moreover, attribute information is naturally integrated into the embeddings since it is the input to the model. As a whole, SEANO is trained in a semi-supervised learning fashion by using both the labeled data and unlabeled data, and the embedding layer, as the bridge between the input layers and output layers, is bound to incorporate all the heterogeneous information in the PLAN.

**Redressing Outliers in Embedding Learning:** Note that the input layers of SEANO include attribute information from both the target node and its neighborhoods. These two sources of inputs are fused at the embedding layer through a weighted sum based on the aggregation weight $\lambda_i$ as shown in Equation 3.1. By incorporating the neighborhoods in the input, SEANO not only collects additional information for embedding learning, but more importantly, also smooths out the noises arising from each individual vertex $i$. Intuitively, if a vertex contains anomalous attributes compared to other vertices in the similar context, SEANO will inherently rely more on neighborhood attributes in order to provide better prediction results. Consequently, it will lead the model towards learning a smaller weight $\lambda_i$ through training. As a result, we can alleviate the negative effect of network outliers through incorporating the neighborhoods information and adaptively learning the aggregation weights. In fact, as we will show in Sec. 3.2.4, the learned weight $\lambda_i$ can be nicely interpreted as outlier score for each vertex $i$ and further used for outlier detection.

**Inductive Embedding Learning:** We point out SEANO can be easily generalized to infer the embedding of a new vertex upon observing its attributes and neighborhoods. This generalization is possible because the embedding $\mathbf{e}_i$ of vertex $i$ is computed using $\mathbf{e}_i = \lambda_i \mathbf{h}^{l_1}(\mathbf{x}_i) + (1-\lambda_i)\mathbf{h}^{l_1}(\overline{\mathbf{x}}_{N_i})$, which merely depends on $\mathbf{x}_i$, $\overline{\mathbf{x}}_{N_i}$, and $\lambda_i$. $\mathbf{x}_i$ and $\overline{\mathbf{x}}_{N_i}$ can be easily obtained when a new vertex $i$ arrives. For $\lambda_i$, we can set it as a constant depending on certain prior knowledge on the normality of new vertices. In this paper, we take a conservative estimation and set $\lambda_i = 0.5$ for unobserved vertices, which we empirically found works well. By doing this, our model can be applied to inductively infer the embeddings of unseen nodes by incorporating the heterogeneous information.

---

[1] This is computed by averaging the attribute vectors of $v_i$'s one-hop neighborhoods in the PLAN.

[2] The activation function in this paper is set as the rectified linear unit, i.e. $\phi(x) = \max(0, x)$.

[3] Note that the operator $+$ indicates element-wise addition.

**3.2.2 Loss Functions** Following the previous definitions, the input values of the softmax layer for label prediction can be formulated as $\mathbf{h}^{l_2}(\lambda_i \mathbf{h}^{l_1}(\mathbf{x}_i) + (1-\lambda_i)\mathbf{h}^{l_1}(\overline{\mathbf{x}}_{N_i}))$, denoted as $\mathbf{h}^{l_1 \oplus l_2}(\mathbf{x}_i, \overline{\mathbf{x}}_{N_i})$ for simplicity. Similarly, the input of the softmax layer for context prediction is $\mathbf{h}^{l_1 \oplus l_3}(\mathbf{x}_i, \overline{\mathbf{x}}_{N_i}) = \mathbf{h}^{l_3}(\lambda_i \mathbf{h}^{l_1}(\mathbf{x}_i) + (1-\lambda_i)\mathbf{h}^{l_1}(\overline{\mathbf{x}}_{N_i}))$.

The supervised learning component of the deep model shown in the left part of Figure 1 is the canonical multilayer perceptron (MLP). Its loss function is:

$$(3.2) \quad L_s = -\sum_{i \in V_L} \log p(y_i | \mathbf{x}_i, \overline{\mathbf{x}}_{N_i})$$

where $p(y_i | \mathbf{x}_i, \overline{\mathbf{x}}_{N_i})$ is the likelihood for the target label and is formally defined as

$$(3.3) \quad p(y_i | \mathbf{x}_i, \overline{\mathbf{x}}_{N_i}) = \frac{\exp\left(\mathbf{h}^{l_1 \oplus l_2}(\mathbf{x}_i, \overline{\mathbf{x}}_{N_i})^T \mathbf{W}^{(s)}_{y_i}\right)}{\sum_{y_j \in \mathcal{Y}} \exp\left(\mathbf{h}^{l_1 \oplus l_2}(\mathbf{x}_i, \overline{\mathbf{x}}_{N_i})^T \mathbf{W}^{(s)}_{y_j}\right)}$$

Here, $\mathcal{Y}$ denotes the set of possible labels. As written in Figure 1, $\mathbf{W}^{(s)}$ is the weight matrix of the softmax layer used in the supervised learning part of the model.

The unsupervised learning component of SEANO is analogous to methods used in Word2Vec [23] and DeepWalk [27]. We adopt the Skip-gram model [23] to capture the relationship between the target vertex (at input) and context vertex (at output). For each node $i$ in a PLAN with attributes $\mathbf{x}_i$, we generate its context $C_i = \{v_{i,1}, v_{i,2}, ..., v_{i,c}\}$. We then construct the loss function as:

$$(3.4) \quad L_u = -\sum_{i \in V} \sum_{v' \in C_i} \log p(v' | \mathbf{x}_i, \overline{\mathbf{x}}_{N_i})$$

where $p(v' | \mathbf{x}_i, \overline{\mathbf{x}}_{N_i})$ is the likelihood of the target context given the attributes of the vertex and its neighborhood:

$$(3.5) \quad p(v' | \mathbf{x}_i, \overline{\mathbf{x}}_{N_i}) = \frac{\exp\left(\mathbf{h}^{l_1 \oplus l_3}(\mathbf{x}_i, \overline{\mathbf{x}}_{N_i})^T \mathbf{W}^{(u)}_{v'}\right)}{\sum_{v \in V} \exp\left(\mathbf{h}^{l_1 \oplus l_3}(\mathbf{x}_i, \overline{\mathbf{x}}_{N_i})^T \mathbf{W}^{(u)}_{v}\right)}$$

Here, $\mathbf{W}^{(u)}$ is the weight matrix of the softmax layer used in the unsupervised part of the model. Note that this formulation is different from the one in DeepWalk [27] because the prediction likelihood $p(v' | \mathbf{x}_i, \overline{\mathbf{x}}_{N_i})$ is conditioned on the attributes $\mathbf{x}_i$ and $\overline{\mathbf{x}}_{N_i}$ instead of the vertex id.

We now discuss how the context $C_i$ is generated for each vertex $i$. We categorize the context of a vertex into the *structure context* and the *label context* extending the ideas presented in [37]. The structure context of a vertex consists of the vertices that are close to the vertex in the network and can be generated through truncated random walks in the network [27]. The label context of a vertex $i$ is defined as the vertices sharing the same label and can be generated by uniformly sampling from vertices with label $y_i$.

For each vertex $i$, we generate in total $c$ context vertices following the steps illustrated in Algorithm 1. Specifically, for a labeled vertex, we generate $c * \alpha$ label context vertices (Line 2-3) and $c * (1-\alpha)$ structure context vertices from the stream of short random walks (Line 4). For an unlabeled vertex, we extract $c$ structure context vertices (Line 6). For generating structure context we essentially follow the same approach as Deepwalk [27].

---
**Algorithm 1** Vertex Context Generation

**Input**: A PLAN $\mathcal{G} = (V, E, X, Y)$, target vertex $i$, size of context $c$, and ratio of label context $\alpha$.
**Output**: Vertex context $C_i$.
1: **if** $i \in V_L$ **then**                    ▷ node $i$ is labeled.
2:     $S_{y_i} = \{j : j \in V_L \land y_j = y_i\}$.   ▷ nodes with label $y_i$.
3:     $C_i \leftarrow$ randomly sample $c * \alpha$ label context from $S_{y_i}$.
4:     $C_i \leftarrow C_i \cup \{$generate $c*(1-\alpha)$ structure context vertices$\}$.
5: **else**
6:     $C_i \leftarrow$ generate $c$ structure context vertices.
7: Return $C_i$.

---

**3.2.3 Model Training** In this part, we discuss how we jointly minimize the supervised loss $L_s$ and the unsupervised loss $L_u$. We start by describing the optimization procedures for each respective part (supervised and unsupervised) of our model. As mentioned previously, the supervised unit in the left part of Figure 1 is the standard MLP. We can easily use back-propagation and gradient descent to train the model [30].

For the unsupervised component with the loss function described in Equation 3.4, training can be rather expensive because it requires iterating through all the vertices in $V$ to compute $\nabla(\log p(v' | \mathbf{x}_i, \overline{\mathbf{x}}_{N_i})$ with Equation 3.5, which is quite inefficient on large datasets. To address this problem, we adopt the negative sampling strategy [23]. Instead of directly working on Equation 3.5 and looking at all the vertices, we replace it with the negative sampling objective proposed in Word2Vec [23]. Applying it to Equation 3.4, we have the following loss function

$$(3.6) \quad \widetilde{L}_u = -\sum_{i \in V} \sum_{v' \in C_i} \Bigg[ \log \sigma \left(\mathbf{h}^{l_1 \oplus l_3}(\mathbf{x}_i, \overline{\mathbf{x}}_{N_i})^T \mathbf{W}^{(u)}_{v'}\right) +$$

**Algorithm 2** Learn Embedding with Outlier Handling
**Input**: A PLAN $\mathcal{G} = (V, E, X, Y)$, $B_1$, $B_2$, and $t$.
**Output**: Embedding $\mathcal{E} = (\mathbf{e}_1, \mathbf{e}_2, ..., \mathbf{e}_n)$ for each vertex $i$.
1: Glorot initialization for weights and biases in the model [8].
2: **while** not converged **and** not reaching max iterations **do**
3:    // *Training on the supervised component.*
4:    Sample $B_1$ labeled instances from $V_L$.
5:    Compute $\nabla L_s$ and update the weights and biases.
6:    // *Training on the unsupervised component.*
7:    $C = \emptyset$, $V_{neg} = \emptyset$.
8:    Sample $B_2$ instances from $V$, denoted as $V_B$.
9:    **for** each node $i \in V_B$ **do**
10:       $C_i \leftarrow$ generate vertex context using Algorithm 1.
11:       $C \leftarrow C \cup C_i$.
12:       **for** each node $v' \in C_i$ **do**    ▷ Negative sampling
13:          $V_{v',neg} \leftarrow$ randomly sample $t$ vertices from $V$.
14:          $V_{neg} \leftarrow V_{neg} \cup V_{v',neg}$.
15:    Compute $\nabla \widetilde{L}_u$ with $C$, $V_{neg}$, and Equation 3.6.
16:    Update weights/biases based on the gradient $\nabla \widetilde{L}_u$.
17: **for** each vertex $i \in V$ **do**
18:    Compute the embedding $\mathbf{e}_i$ using Equation 3.1.
19: Return $\mathcal{E} = (\mathbf{e}_1, \mathbf{e}_2, ..., \mathbf{e}_n)$.

$$\sum_{\tilde{v} \in V_{v',neg}} \log \sigma \left( -\mathbf{h}^{l_1 \oplus l_3}(\mathbf{x}_i, \overline{\mathbf{x}}_{N_i})^T \mathbf{W}_{\tilde{v}}^{(u)} \right) \Bigg]$$

where $\sigma$ is the Sigmoid function, and $V_{v',neg}$ is a negative set with $t$ randomly selected negative samples.

To jointly minimize the supervised loss and unsupervised loss in our model, we use Mini-Batch Stochastic Gradient Descent and alternate updating the parameters between the two components of the model. As illustrated in Algorithm 2, we jointly train the two components by alternating between them with a batch size of $B_1$ on the supervised part (Line 4-5) and $B_2$ on the unsupervised part (Line 7-16). Note that these two components are tightly connected as they share the first $l_1 + 1$ layers in the neural network (see Figure 1). As a result, both the supervised component and unsupervised component will update the parameters in the shared layers. In addition, to restrict $\lambda_i$ in the range $[0,1]$, we set $\lambda_i = \sigma(\omega_i)$ and optimize over $\omega_i$. After training the model, the embedding for each node is composed of the activation values at the embedding layer, as computed by Equation 3.1.

#### 3.2.4 Outlier Detection Using SEANO
As shown in Figure 1, SEANO takes the inputs from vertex attributes $\mathbf{x}_i$ and its neighborhood attributes $\overline{\mathbf{x}}_{N_i}$. These two are fused into the embedding layer after going through a series of hidden layers. The fusion of the two sources of information depends on the aggregation weights $\lambda_i$, which are learned through the model training phase. In the case where the vertex is anomalous (its attributes are very different from other vertices in the same context), SEANO will learn to downplay the input from the vertex attributes $\mathbf{x}_i$ and depend more on neighborhood attributes $\overline{\mathbf{x}}_{N_i}$ for performing the predictions, as motivated by the potential decrease of loss function. Arguably, this design smooths out noises from an individual vertex and can adaptively improve the robustness of the embeddings. More importantly, we point out that the weight parameter $\lambda_i$ can be interpreted as the outlier score for each vertex $i$ in the PLAN. After training our model, each vertex $i$ in the PLAN is associated with a weight parameter $\lambda_i$, which always lies in the range $[0, 1]$. A low value of $\lambda_i$ indicates that the attributes of vertex $i$ are not informative at predicting class label and graph context as compared to the majority of vertices. This is a strong signal that its attributes, label, and graph structure do not conform to the underlying pattern in the PLAN. Following this idea, we interpret $\lambda_i$ as the outlier score for vertex $i$. The lower is the outlier score, the more likely is the vertex to be an outlier.

| Dataset | # Classes | # Attr. | # Nodes | # Edges |
|---|---|---|---|---|
| Cora | 7 | 1,433 | 2,708 | 5,429 |
| Citeseer | 6 | 3,703 | 3,327 | 4,732 |
| Pubmed | 3 | 500 | 19,717 | 44,338 |
| Houston | 2 | 3 | 39,215 | 155,638 |
| Houston-large (high-res) | 2 | 3 | 3,926,150 | 15,692,353 |

Table 3: Dataset Information

### 4 Experiments and Analyses
**Datasets:** The datasets we use in our experiments are described in Table 3. The first three datasets, Cora, Citeseer, and Pubmed[4], were used in prior work [31, 37], where vertices represent published papers and edges (undirected) denote the citations among them. Each paper contains a list of keywords and they are treated as attributes. The papers are classified into multiple categories based on the topics and this information serves as labels. The other two datasets are constructed from satellite images with pixels as vertices and sensor data as real-value attributes (see Section 4.4 for more details on PLAN construction).

To evaluate robustness of proposed methods to outliers, we randomly select 5% of the vertices in the training dataset, including both labeled and unlabeled data, and modify their attributes following the natural perturbation scheme described by Song *et. al* [32, 19]. Specifically, for a selected node $i$ for noise injection, we randomly pick another $m = \min(100, \frac{n}{4})$ vertices from the PLAN and select the node $j$ with the most different attributes from node $i$ among the $m$ nodes, i.e., maximizing $\|\mathbf{x}_i - \mathbf{x}_j\|_2$. We then replace the attributes $\mathbf{x}_i$ of node $i$ by $\mathbf{x}_j$.

**Baseline Methods:** In this experiment, we evaluate the embedding methods on the task of vertex classification. We compare the performance of SEANO with several groups of baselines as follows.

[4] http://linqs.cs.umd.edu/projects/projects/lbc

- SVM [33], TSVM [14], and Doc2Vec [16]: This group of baselines primarily use the vertex attributes but not the graph structure of the PLAN. SVM (Support Vector Machine) classifier is trained based on the attribute information and the labeled data. TSVM (Transductive Support Vector Machine) is a variant of the regular SVM that uses both labeled data and unlabeled data for training (testing data is incorporated in the unlabeled data). Doc2Vec treats each node as a document and learns the distributed representation of each node, which is further used to train an SVM classifier.

- Node2Vec [9], Node2Vec+, and TADW [36]: This group of baselines do not leverage the label information during embedding learning process. Node2Vec is an improved version of DeepWalk [27], where it generates embedding of each vertex based on the graph structure. We additionally concatenate the Node2Vec embeddings with original vertex attributes and we call this baseline, Node2Vec+. TADW uses a more systematic way to combine the topological information and attributes into the embeddings following a similar idea to DeepWalk. To use these different embeddings for classification, we train an SVM classifier on the labeled data for each of them.

- TriDNR [26], Planetoid [37], CNN-Cheby [5], GCN [15], GraphSAGE [10]: This group of baselines are competitive strawmen in that all of them incorporate information on attributes, structure, and labels of the PLAN. All of them are neural-network-based methods though they have different designs of network architecture and use different objective functions for training. Among them, Planetoid contains two versions: one for transductive learning (Planetoid-T) and the other for inductive learning (Planetoid-I). TriDNR, CNN-Cheby, and GCN can only work on transductive learning while GraphSAGE is designed for inductive learning.

- SEANO-0.5, SEANO-1.0: These are two variants of SEANO with constant $\lambda_i$ value. While SEANO uses parameters $\lambda_i$ at the embedding layer for aggregation and learns the parameters through training, these two variants statically fix $\lambda_i$ to 0.5 and 1.0 respectively. SEANO-0.5 assumes vertex attributes and neighborhood attributes contribute equally to the embeddings while SEANO-1.0 entirely ignores neighborhood attributes from the inputs.

**Experimental Setup:** We follow the same data split strategy as previous work [37, 15]. Specifically, we randomly select 20 instances from each class and treat them as the labeled data for training. We randomly sample 1,000 of the remaining data as the validation dataset for the purpose of parameter tuning. We sample another 1,000 from the rest of data and treated them as the testing dataset for evaluation.

All the experiments were conducted on a machine running Linux with an Intel Xeon E5-2680 CPU (28 cores, 2.40GHz) and 128GB of RAM. We implement SEANO using the TensorFlow package in Python. The code is publicly available at http://jiongqianliang.com/SEANO. We use the SVM-light package for SVM and TSVM (with RBF kernel)[5]. For other baselines, we adapt the source code from the original authors. The dimensionality of the embedding $r$ is set to 50 for all the methods wherever applicable. For SEANO, we additionally set $c = 8$, $t = 6$ (as recommended by [27, 23]), and $l_1 = l_2 = l_3 = 1$[6]. For other hyper-parameters in our model ($B_1$, $B_2$, and $\alpha$) and other baselines, we try our best to tune them for the best performance using the validation dataset.

**4.1 Transductive Learning** In this experiment, we conduct network embedding in the PLAN in a transductive manner (testing data accessible during training) and evaluate different embedding methods on the task of node classification. We run this experiment on the first four datasets as well as their noisy versions[7]. The classification accuracy of all the compared methods on these datasets is reported in Table 4. We highlight the following main observations:

**1) Vertex attributes, graph structure, and label information are all useful for improving the quality of the embeddings.** As we incorporate more information for embedding learning (from top to bottom in Table 4), we tend to obtain better embeddings, which is shown by the improvement on the classification performance. If we compare the methods in the first two groups with the ones in the last two groups, we can clearly observe the significant performance gap. Note the main difference between the first two groups with the last two groups is that the former merely use one or two types of information for embedding learning, while the latter leverage all three types of information (vertex attributes, graph structure, and partially available labels). This observation is more obvious on the Cora dataset, where SEANO, GCN $\gg$ Node2Vec+ $\gg$ Node2Vec, SVM.

**2) Heterogeneous information needs to be fused in a systematic manner in order to achieve quality embedding.** Simple strategies, such as concatenating network embeddings with attributes, do not al-ways work. This can be observed by comparing the performance of Node2Vec+ against the simple SVM. Node2Vec+ is different from the SVM in that it extends the original vertex attributes by concatenating the vertex embeddings. Yet we observe that Node2Vec+ performs worse than the SVM on Citeseer, Pubmed and Houston datasets. The relatively poor performance of Node2Vec+ compared to more advanced methods, such as Planetoid-T and GCN, indicates that network structure is helpful in improving embedding quality only when it is incorporated in a principled way. This validates the necessity for jointly learning the embeddings of a PLAN.

---
[5]http://svmlight.joachims.org/
[6]While increasing these values slightly improves the performance, it is much more computationally expensive.
[7]Majority of the baselines cannot scale to Houston-large dataset and we show its results in the case study section instead.

|  | Cora | Citeseer | Pubmed | Houston | Cora* | Citeseer* | Pubmed* | Houston* |
|---|---|---|---|---|---|---|---|---|
| SVM | 58.3 | 58.0 | 73.5 | 96.0 | 58.1 | 58.0 | 73.5 | 96.0 |
| TSVM | 58.3 | 62.3 | 66.9 | N/A | 55.6 | 61.4 | 66.5 | N/A |
| Doc2Vec | 41.5 | 41.2 | 62.0 | N/A | 39.0 | 37.2 | 62.9 | N/A |
| Node2Vec | 65.2 | 41.7 | 61.4 | 66.7 | 65.2 | 41.7 | 61.4 | 66.7 |
| Node2Vec+ | 68.5 | 53.1 | 62.2 | 75.3 | 68.3 | 53.1 | 62.2 | 75.3 |
| TADW | 64.4 | 56.3 | 48.7 | 83.7 | 59.8 | 48.8 | 46.5 | 83.7 |
| TriDNR | 59.9 | 43.2 | 68.5 | N/A | 45.6 | 39.7 | 68.3 | N/A |
| Planetoid-T | 75.7 | 62.9 | 75.7 | 96.7 | 66.9 | 58.3 | 72.7 | 94.5 |
| CNN-Cheby | 79.2 | 68.1 | 75.3 | 92.9 | 79.0 | 67.0 | 74.4 | 92.1 |
| GCN | 81.5 | 72.1 | 79.0 | 93.9 | 81.0 | 70.2 | 76.5 | 92.8 |
| SEANO-0.5 | 81.4 | 69.8 | 71.6 | 97.1 | 80.6 | 69.0 | 66.8 | 96.8 |
| SEANO-1.0 | 76.9 | 68.9 | 68.7 | 96.8 | 75.6 | 66.8 | 63.0 | 96.3 |
| SEANO | **82.0** | **74.3** | **79.7** | **97.3** | **81.6** | **73.0** | **77.4** | **97.0** |

Table 4: Classification accuracy (in percentage) of using different transductive methods on the original datasets and the noisy datasets (mark with ∗). "N/A" of TriDNR and Doc2Vec means the methods are not applicable on datasets with non-binary attributes. "N/A" of TSVM indicates it cannot finish training in 24 hours.

**3) SEANO generates the best-quality embeddings and consistently outperforms other methods on vertex classification.** In particular, it consistently outperforms the state-of-the-art methods, including Planetoid-T, CNN-Cheby, and GCN. We argue that one major reason for the performance lift is because SEANO is able to redress the adverse effect of the network outliers during the embedding learning phase. One evidence for it is that SEANO performs better than its variants SEANO-0.5 and SEANO-1.0, which use exactly the same neural network architecture except fixing the aggregation weights $\lambda_i$. This reveals that using adaptive aggregation weights based on the outlierness of the vertices (SEANO) has an obvious advantage over the alternatives, which either ignore neighborhood attributes (SEANO-1.0) or evenly merge the input signals from vertex attributes and neighborhood attributes (SEANO-0.5). We should also emphasize that SEANO shows the least performance drop when applied to the noisy datasets especially in contrast to the competitive strawmen (GCN, SEANO-0.5, SEANO-1.0).

|  | Cora | Citeseer | Pubmed | Houston |
|---|---|---|---|---|
| SVM | 58.3 | 58.0 | 73.5 | 96.0 |
| Planetoid-I | 61.2 | 64.7 | 77.2 | 96.2 |
| GraphSAGE | 60.0 | 58.5 | 73.0 | 93.6 |
| SEANO | **80.2** | **72.8** | **78.9** | **97.2** |

Table 5: Classification accuracy (in percentage) of different inductive methods on the four datasets.

**4.2 Inductive Embedding Learning** As we discussed in Section 3.2, SEANO is also designed to support inductive embedding learning. In this experiment, we show that SEANO is able to infer quality embeddings for vertices that are unobserved during model training. Inductive learning is typically more challenging and several of the previous baselines cannot be applied to this setting. For comparisons, we adopt the inductive variant of Planetoid (Planetoid-I) [37] and state-of-the-art method GraphSAGE [10]. For the convenience of reference, we still keep SVM as a baseline though it performs exactly the same with the previous experiment. For the purpose of inductive learning, the testing dataset with 1,000 vertices is held-out and cannot be accessed during the training phase. The remaining of the experimental settings are similar to those in the transductive learning experiment. Table 5 shows the performance of SEANO in inductive learning compared to other methods.

As shown in Table 5, SEANO performs significantly better than other methods. The largest gap lies in Cora and Citeseer dataset, where we respectively observe an improvement on the accuracy of 19% and 8% over the second best baseline. By comparing the performance of SEANO in Table 5 with the one in Table 4, we can see that SEANO performs almost equally well on the inductive learning with only a slight decrease of accuracy. The results of this experiment demonstrate that even when the testing dataset is not observed, SEANO is still able to learn the embeddings reasonably well, **significantly** outperforming the state-of-the-art.

|         | Cora* | Citeseer* | Pubmed* | Houston* |
|---------|-------|-----------|---------|----------|
| Attr.-only | 17.0 | 16.3 | 16.3 | 15.1 |
| Planetoid-T | 5.9 | 31.3 | 4.4 | 30.6 |
| GCN | 1.4 | 1.2 | 5.1 | 1.9 |
| AMEN | 7.4 | 10.2 | 6.1 | N/A |
| ALAD | 36.3 | **55.4** | 43.4 | N/A |
| SEANO-embed | 6.7 | 16.9 | 48.4 | 10.9 |
| SEANO | **41.5** | 54.2 | **49.8** | **47.4** |

Table 6: Outlier detection performance comparison. Precision is reported as measurements (in percentage). "N/A" means the method cannot finish in 24 hours or runs out of memory.

**4.3 Outlier Detection using SEANO** Besides learning robust embeddings for PLANs, here we show that SEANO is also capable of detecting network outliers by interpreting the aggregation weights $\lambda_i$ as outlier scores. We compare the performance of SEANO on flagging the injected network outliers [32] against the following baselines: 1) Attribute only method (Attr.-only), which runs Isolation Forest [20] on the vertex attributes for outlier detection. 2) Planetoid-T, GCN, and SEANO-embed, which apply Isolation Forest algorithm on the network embeddings generated by Planetoid-T, GCN, and SEANO respectively. Note that these methods generate best-quality embeddings as shown in the transductive learning experiment. 4) AMEN [28] and ALAD [21] which are state-of-the-art attributed networks outlier detection algorithms. We run all the methods on the noisy datasets used above, where 5% outliers are injected using the natural perturbation scheme [32]. We set the number of outliers to detect as the number of injected outliers, and compute the precision. Table 6 shows the performance on outlier detection of different methods.

We see clearly that SEANO performs reasonably well on this challenging task with performance comparable to state-of-the-art methods that are specifically designed for detecting attributed network outliers. By leveraging the aggregation weights in the model, SEANO comfortably outperforms the best baseline ALAD (designed for attributed network outlier detection) on Cora∗ and Pubmed∗ while falling slightly short on Citeseer∗. This is impressive performance considering that SEANO is not designed for network outlier detection in the first place. SEANO also dominates other embedding-based strawmen (GCN, Planetoid-T, SEANO-embed) – many of which perform very poorly on this task. We conjecture this is because the embeddings integrate all the information into a coherent vector representation and cannot distinguish information of attributes (served as outlier indicator features) with label and graph information (contextual features) [21, 19].

**4.4 Case Study: Flood Mapping** We examine the scalability and effectiveness of SEANO on a challenging real-world problem (Flood Mapping). To this end, we use a high-resolution satellite image of Houston collected immediately after the 2016 Houston flood using synthetic aperture radar. There are two raw attributes (HH and HV) from radar and another representing geographic elevation for each pixel in the image. HH and HV measure the polarity of waves reflected by a material and are helpful in distinguishing water from land. The goal is to conduct semi-supervised learning to discriminate water from land. We convert the image into an undirected graph following the approach proposed by Cour *et al.* [4]. Each pixel of the image is treated as one vertex and has edges to nearby pixels within a Euclidean distance of 1.5 units. The three attributes of each vertex (HH, HV, elevation) are used as vertex attributes. The resulting PLAN, denoted as Houston-large, comprises 3,926,150 vertices and 15,692,353 edges (a down-scaled low-resolution version, called Houston, is used in the previous experiments). The ground-truth label (water or land) is provided by domain experts, from which we sample 100 instances as labeled data for training.

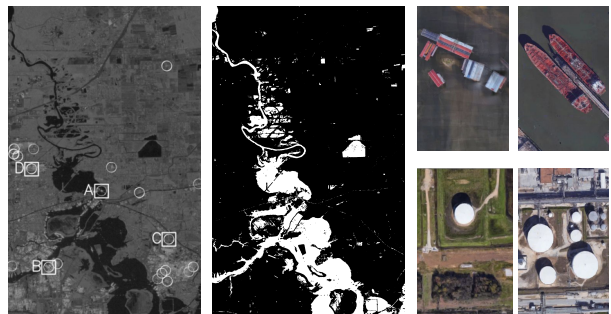

(a) SAR image   (b) Results   (c) Outliers

Figure 2: Visualization. (a) SAR image of the Houston area. (b) Water delineation results of using SEANO. (c) Google Maps of the area surrounding 4 representative outliers (from top to bottom, left to right, they correspond to A, B, C and D in (a)).

We run different methods on Houston-large in the similar setting as the transductive learning experiment (results in Table 7). Here we include baselines that are specialized in flood mapping and image segmentation from the remote-sensing and computer vision community in addition to the top three algorithms from our previous analysis (SEANO, Planetoid-T and SVM). Hug-fm is a state-of-the-art algorithm for semi-supervised water delineation on satellite images (supervised by domain expert) [18]. Norm-thr [22] is a modern split-based automatic thresholding method for water delineation developed in the remote sensing community. Otsu [25] is a venerable clustering-based thresholding method widely used in computer vision and remote sensing. Watershed (W.s.) [2] algorithm is a region-growing technique based on the labeled markers. It is clear from Table 7 that SEANO comfortably outperforms the baselines on classifying flooded areas. It is also worth noting that many of the competitive baselines (e.g. SVM, Planetoid) that worked reasonably well on the low-resolution Houston data, are not as effective in this setting (compared with Table 4). Finally, we note that even small (1%) statistically significant improvements in $F_1$ score can yield significant savings in large urban settings (billions of dollars) and improved prioritization of post-disaster emergency relief efforts [22, 18].

|  | SEANO | Planetoid | SVM | Hug-fm | Norm-thr | Otsu | W.s. |
|---|---|---|---|---|---|---|---|
| *Accuracy* | **96.9** | 94.5 | 94.5 | 95.8 | 95.4 | 89.8 | 89.0 |
| $F_1$ *score* | **89.9** | 84.1 | 83.8 | 86.8 | 83.7 | 73.9 | 68.0 |

Table 7: Water/land classification performance (in percentage).

We also visualize the original data and the results output by SEANO in Figure 2. Comparing the water delineation result in Figure 2b with the original satellite image in Figure 2a, we can observe that SEANO accurately delineates water areas of different shapes (e.g. long thin rivers). In addition, the white circles in Figure 2a are the most anomalous vertices based on the outlier scores output by SEANO. Visually, most of the outliers are very bright and dazzling compared to surrounding pixels. To find out what those outliers are in reality, we cross-reference them with Google Maps. Figure 2c shows the areas associated with 4 representative outliers in Google Maps. The first row in Figure 2c shows shipping containers and ships in water bodies (corresponding to *A* and *B* in Figure 2a) while the second row presents areas with large white cylindrical tanks (often housing treated water) that are common in factories (*C* and *D* in Figure 2a). We look into the top 80 outliers and find that a majority of them fall into these cases. A common factor is that they contain strongly reflective metal surfaces, causing them to have higher HH and HV values in contrast to neighboring pixels. To summarize, this case study demonstrates that SEANO is efficient (can scale to large problems) and effective (when compared to the state-of-the-art), on a real-world problem with outlier-effects.

## 5 Conclusions

We propose a semi-supervised inductive learning framework to learn robust embeddings that jointly preserve graph proximity, attribute affinity and label information while accounting for outlier effects. We extend the proposed model for detecting network outliers. Our experiments on real-world data and a case study on flood mapping demonstrate the efficacy of our method over the state of the art. As future work, we plan to adapt SEANO to other types of networks, such as attributed hypernetworks and heterogeneous information networks.

**Acknowledgments.** This work is supported by NSF of the United States under grant EAR-1520870, DMS-1418265 and CCF-1645599. All content represents the opinion of the authors, which is not necessarily shared or endorsed by their sponsors.